\documentstyle[11pt,verynewpasp,twoside,psfig,epsf]{article}
\def\edcomment#1{\iffalse\marginpar{\raggedright\sl#1\/}\else\relax\fi}
\reversemarginpar

\begin{document}
\title{Radio recombination lines from starburst galaxies : high and low density
ionized gas}
\author{Niruj R. Mohan}
\affil{Raman Research Institute, Bangalore 560080, India \& \\
Indian Institute of Science, Bangalore 560012, India\\}
\author{K.R. Anantharamaiah}
\affil{Raman Research Institute, Bangalore 560080, India}
\author{W.M. Goss}
\affil{NRAO, Socorro, NM 87801, USA}
\begin{abstract}
Radio recombination lines (RRL) at 8 GHz and 15 GHz detected from 
four starburst galaxies
are shown to arise in compact high density HII regions, which are 
undetectable below $\sim$4 GHz. Detection of an RRL at 1.4 GHz towards
one galaxy and upper limits in the other three are consistent with the
presence of an equal amount of low density diffuse gas. Continuum 
flux density measurements using the GMRT will be 
important in constraining the properties of the diffuse gas.
\end{abstract}
\vspace{-0.8cm}
\section{Introduction}
\vspace{-0.3cm}
RRL and radio continuum studies of nuclear starbursts in galaxies
are proving to be useful not only because of the absence of extinction
but also because different density components of the ionized gas 
can be accessed through
observations of RRLs at different frequencies (Zhao et al. 1996).  
We report a multi-frequency RRL and continuum study of four starburst galaxies.
\vspace{-0.5cm}
\section{Observations and Modeling}
\vspace{-0.3cm}
We have observed RRLs at 1.4 GHz, 8.3 GHz and 15 GHz using the 
VLA{\footnote {The National Radio Astronomy Observatory is a facility of
the National Science Foundation operated under cooperative agreement
by Associated Universities, Inc.}}
from four starburst galaxies : NGC 253, NGC 3628, and NGC 3690 and
IC694 in the Arp 299 system. The two higher frequency lines were detected
in all four galaxies and the line at 1.4 GHz was detected only in NGC 253.
The line emission arises in the central 
5$-$8$^{\prime\prime}$~nuclear region in these galaxies.
The peak line strengths for NGC 253 are about 5-8 mJy.
The 8 and 15 GHz lines are typically 0.5$-$1.0 mJy in the other three
galaxies and the 3$\sigma$~upper limits to the 1.4 GHz lines are $\sim$1 mJy.
We model the ionized gas as a collection of spherical uniform HII
regions characterised by a single temperature, density and size. 
The total number of such
HII regions is obtained by comparing the computed continuum and line
emission for each model with the observed values. Other derived parameters
are the flux of ionizing photons and the mass of the gas.
Valid solutions are chosen based on constraints which are described 
in Anantharamaiah et al (1993).
\begin{figure}
\plotfiddle{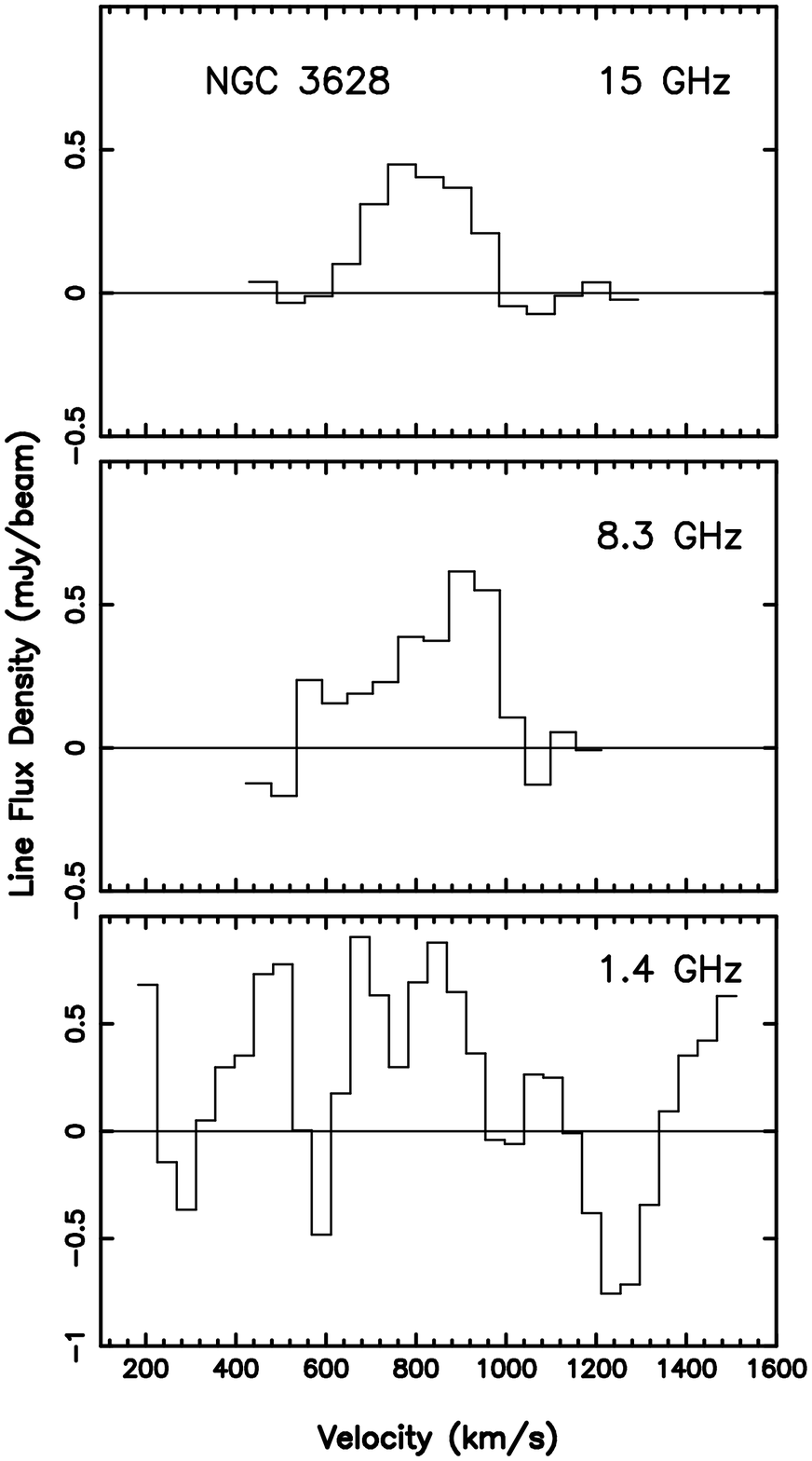}{0.5in}{0}{21}{21}{-180}{-115}
\plotfiddle{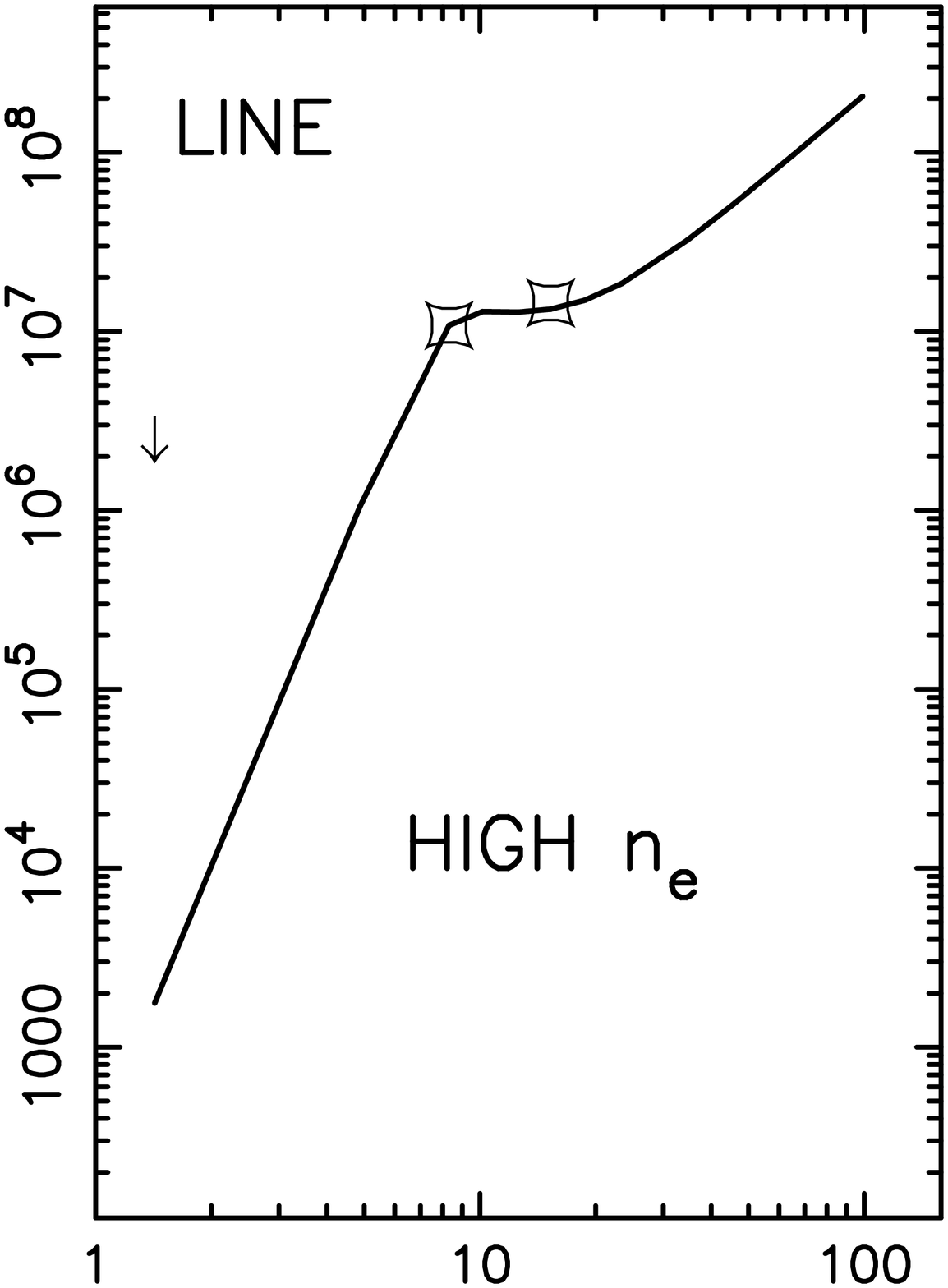}{0.5in}{0}{15}{10}{-50}{25}
\plotfiddle{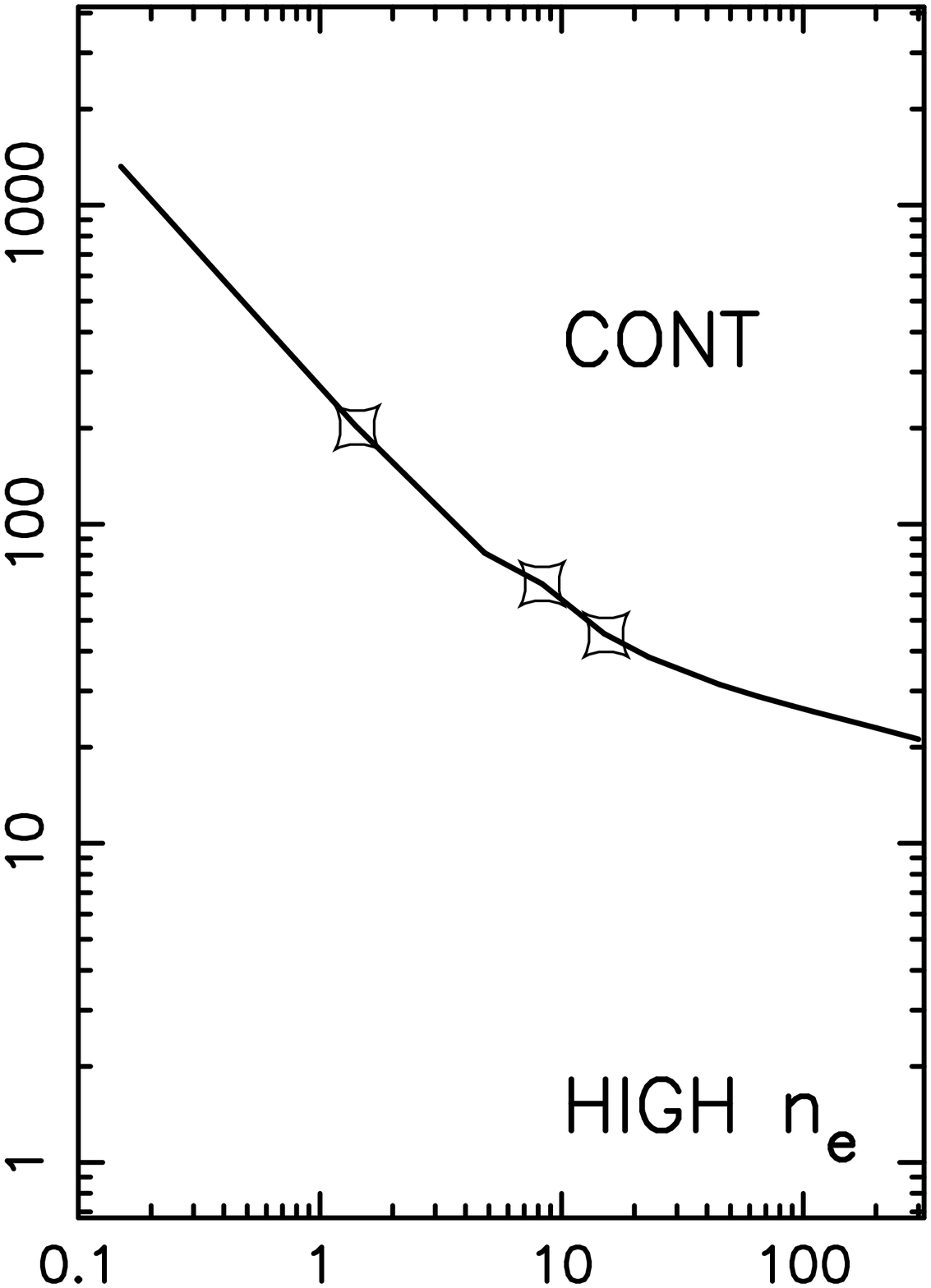}{0.5in}{0}{15}{10}{50}{75}
\plotfiddle{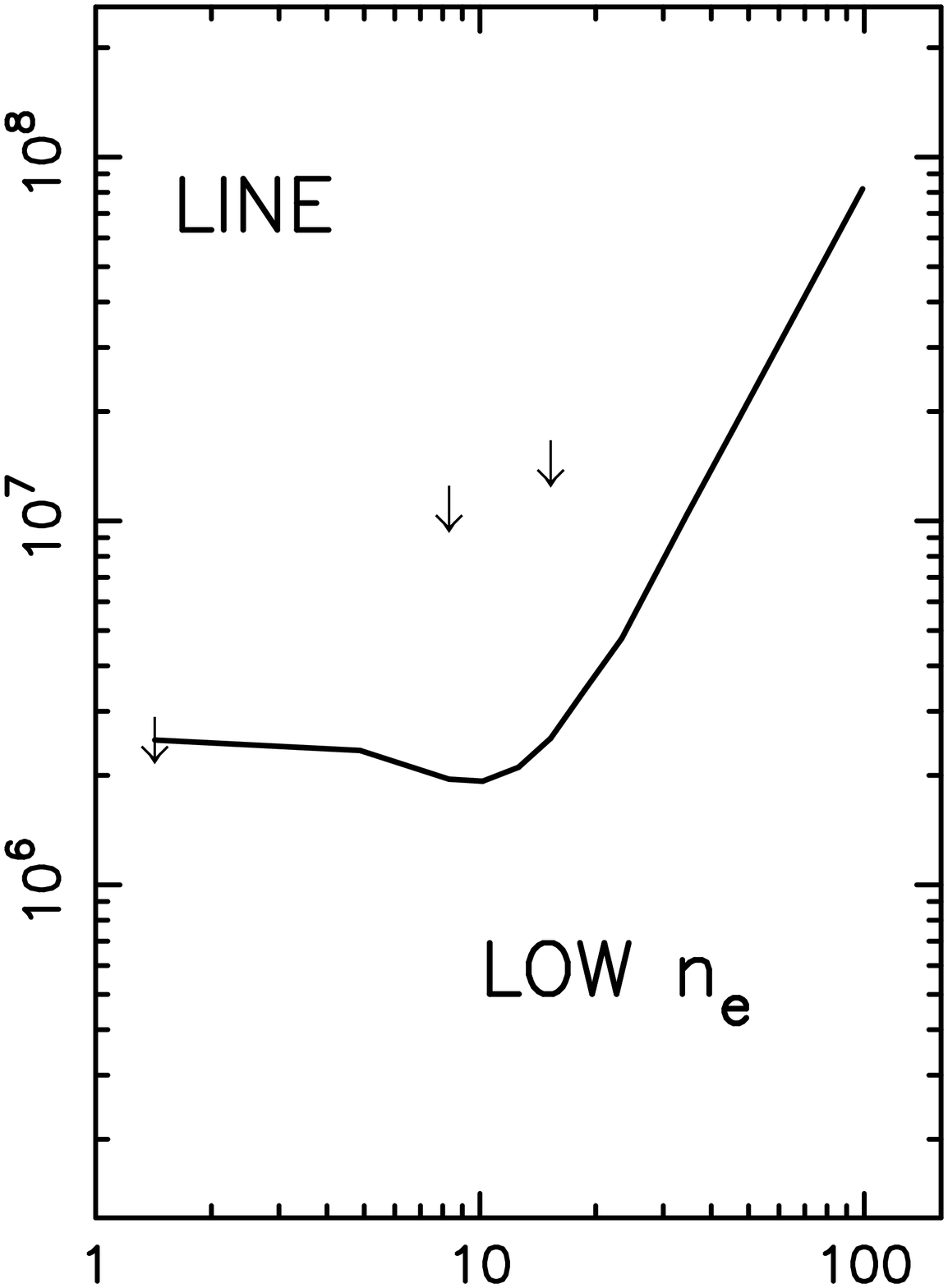}{0.5in}{0}{15}{10}{-50}{40}
\plotfiddle{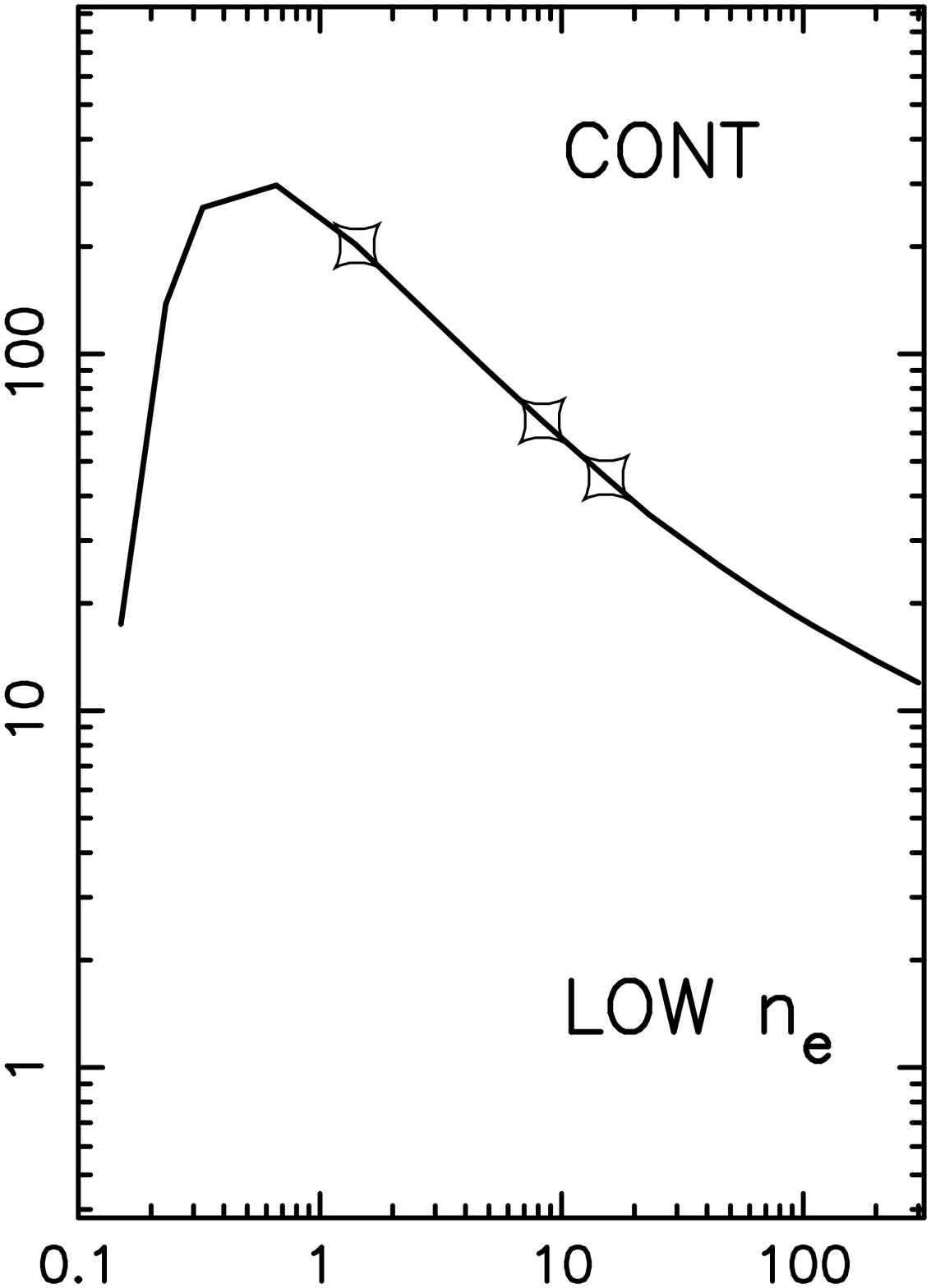}{0.5in}{0}{15}{10}{50}{90}
\vspace{-1.4in}
\caption{{\it Left: } The 2 cm, 3.6 cm and 20 cm (20 cm is a 
non-detection) RRL spectra towards 
NGC 3628. {\it Right: }The model predictions for the line and
continuum flux densities for the high density (T$_e$=10000 K, n$_e$=10$^4$~cm
$^{-3}$, l=1 pc) and low density (T$_e$=10000 K, n$_e$=100 cm$^{-3}$, l=5 pc)
gas versus frequency (in GHz) in NGC 3628. The vertical scales are 
in units of mJy-Hz and mJy for the line and the continuum respectively.
The results are similar for the other galaxies.}
\vspace{-0.6cm}
\end{figure}
\vspace{-0.6cm}
\section{Properties of the Ionized Gas and Conclusions}
\vspace{-0.3cm}
The models show that the 8.3 GHz and 15 GHz lines originate in 
a population of compact
(0.1$-$5 pc) high density (5000$-$50000 cm$^{-3}$) HII regions
with low total volume filling factor ($<$~10$^{-4}$). These
lines arise from internal emission in the HII regions.
Since the NIR and optical data imply much lower densities, this
component is probably not detected in these bands due to high extinction.
In all four galaxies, the photon flux neccesary for ionizing 
this gas is equal to or greater than that derived using 
conventional means. This result could 
lead to an upward revision of their star formation rates. 

The above gas is practically undetectable at frequencies 
\raisebox{-0.8ex}{$\stackrel{<}{\sim}$}4 GHz.
Modeling the detected H166$\alpha$~line at 1.4 GHz from NGC 253 
indicates that this line arises from 
low density (10$-$100 cm$^{-3}$) diffuse (5$-$100 pc) 
HII regions with an area filling factor $>$~0.1. The upper limits
to the 1.4 GHz line for the other three galaxies give very similar
results. Although the detection of lines from this 
component in the other three galaxies might be 
difficult with current instruments, measuring the 
continuum flux densities at $\nu<$1 GHz from the nuclear region
using the GMRT will enable us to strongly constrain 
the properties of this gas.

These observations are a first step towards deriving properties 
of the ionized gas at different densities in starburst regions 
and hence towards studying the star formation at different 
time scales since the lifetime of an HII region depends on its density.

\vspace{-0.1cm}


\vspace{-0.6cm}
\begin{references} 
\vspace{-0.5cm}
\reference Anantharamaiah, K. R., Zhao, J. H., Goss, W. M., \& Viallefond, F. 1993, 
\apj, 419, 585
\reference Zhao, J. H., Anantharamaiah, K. R., Goss, W. M., \& Viallefond, F. 1996, 
\apj, 472, 54
\end{references}
\end{document}